\documentclass[12pt]{article}
\usepackage{amsmath}
\usepackage{graphicx,psfrag,epsf}
\usepackage{enumerate}
\usepackage{natbib}
\usepackage{url} 

\usepackage{graphicx}
\usepackage{slashbox}
\usepackage{amssymb}
\usepackage[english]{babel}
\usepackage{amssymb,amsmath,multirow,array}
\usepackage[english]{babel}
\newtheorem{theorem}{Theorem}

\newtheorem{remark}{Remark}

\newcommand{\blind}{0}

\date{}
\begin{document}


\def\spacingset#1{\renewcommand{\baselinestretch}%
{#1}\small\normalsize} \spacingset{1}


  \title{Zero-inflated generalized extreme value regression model for binary data and application in health study}
  \author{\textbf{DIOP Aba}\footnote{Corresponding author}\\
  \textit{E-mail: aba.diop@uadb.edu.sn} \\
  \textit{Equipe de Recherche en Statistique et Modèles Aléatoires} \\
  \textit{Université Alioune Diop, Bambey, Sénégal} \\
  \textbf{DEME El Hadji} \\
  \textit{E-mail: elhadjidemeufrsat@gmail.com} \\
  \textit{Laboratoire d'Etude et de Recherche en Statistique et Développement} \\
  \textit{Université Gaston Berger, Saint-Louis, Sénégal} \\ 
  \textbf{DIOP Aliou}\\
 \textit{E-mail: aliou.diop@ugb.edu.sn} \\
 \textit{Laboratoire d'Etude et de Recherche en Statistique et Développement} \\
  \textit{Université Gaston Berger, Saint-Louis, Sénégal}
}  
\maketitle
\if0\blind

\bigskip
\begin{abstract}
Logistic regression model is widely used in many studies to investigate the relationship between a binary response variable $Y$ and a set of potential predictors $\mathbf X$. The binary response may represent, for example, the occurrence of some outcome of interest ($Y=1$ if the outcome occurred and $Y=0$ otherwise). When the dependent variable $Y$ represents a rare event, the logistic regression model shows relevant drawbacks. In order to overcome these drawbacks we propose the Generalized Extreme Value (GEV) regression model. In particular, we suggest the quantile function of the GEV distribution as link function, so our attention is focused on the tail of the response curve for values close to one. A sample of observations is said to contain a cure fraction when a proportion of the study subjects (the so-called cured
individuals, as opposed to the susceptibles) cannot experience the outcome of interest. One problem arising then is that it is usually unknown who are the cured and the susceptible subjects, unless the outcome of interest has been observed. In these settings, a logistic regression analysis of the relationship between $\mathbf X$ and $Y$ among the susceptibles is no more straightforward. We develop a maximum likelihood estimation procedure for this problem, based on the joint modeling of the binary response of interest and the cure status. We investigate the identifiability of the resulting model. Then, we
conduct a simulation study to investigate its finite-sample behavior, and application to real data.
\end{abstract}

\noindent%

\spacingset{1.45} 
\section{Introduction}
\label{sec:intro}
Binary regression model is an important special case of the generalized linear models (see \cite{mnelder89} and \cite{hl2000}). Currently, the logistic regression model, with its convenient interpretation and implementation, has been routinely employed to estimate and predict the probability of being infected in medical studies. Estimating a regression model with a cure fraction can be viewed as a zero-inflated regression problem. Zero-inflation occurs in the analysis of count data when the observations contain more zeros than expected (see \cite{lambert92} and \cite{famoy06}). Zero-inflated binomial (ZIB) models have been proposed by \cite{h2000} and \cite{ddd11}. Failure to account for these extra zeros is known to result in biased parameter estimates and inferences. The problem of estimating a logistic regression model from binary response data with a cure fraction, when the cure probability is modeled by a logistic regression were study by \cite{ddd11}. They first investigate the identifiability question in this model. Then, the proposed estimator is obtained by maximizing the joint likelihood for the binary response of interest and the cure indicator. They prove the almost sure asymptotic existence, the consistency, and the asymptotic normality of this estimator. \\

When the logistic regression model is employed, it is assumed that the response curve between the covariates and the probability is symmetric. This assumption may not always be true, and it may be severely violated when the number of observations in the two response categories are significantly different from each other. It is well known that standard algorithms such as logistic regression do not perform well in this setting as they tend to underestimate the probability of the rare class.

This unbalance is not uncommon when we consider binary rare events data (i.e. binary dependent variables with a very
small number of ones) which happens with only a small probability. Applying a nonflexible link function to the data with this special feature may result in link misspecification. Consequences of link misspecification have been studied by a number of authors in the literature (see, for example \cite{czado92} and \cite{king01}). In particular, for independent binary observations, \cite{czado92} show that falsely assuming a logistic link leads to a substantial increase in the bias and mean squared error of the parameter estimates as well as the predicted probabilities, both asymptotically and in finite samples.

In GLM literature (see \cite{mnelder89}, \cite{dob08} and \cite{ag02}) several models for binary response variable have been proposed by considering different link functions: logit, probit, log-log (quantile function of the Gumbel random variable) and complementary log-log models. However, the most used model for binary variables is the logistic regression. The logistic regression shows same important drawbacks in rare events studies: the probability of rare event is underestimated and the logit link is a symmetric function, so the response curve approaches zero as the same rate it approaches one. Moreover, commonly used data collection strategies are inefficient for rare event data (King Zeng, 2001). The bias of the maximum likelihood estimators of logistic regression parameters in small sample sizes, that has been well analysed in literature (\cite{mnelder89}, \cite{ml77} and \cite{mmc85}), is amplified in the rare event study. Most of these problems are relatively unexplored by literature (\cite{king01}).\\

The main aim of this paper is to overcome the drawbacks of the logistic regression in rare events studies by proposing a new model for binary dependent data with an asymmetric link function given by the quantile function of the Generalize Extreme Value (GEV) random variable for the infected model and a logit link function for the cure fraction. In the extreme value theory, the GEV distribution is used to model the tail of a distribution (\cite{k00} and \cite{c01}). Since we focus our attention on the tail of the response curve for the values close to one, we have chosen the GEV distribution.

The rest of this paper is organized as follows. In Section \ref{sec:model}, we describe the problem of GEV regression model with a cure fraction, and we propose an estimation method adapted to this setting. The proposed procedure is based on a joint regression model for the binary response of interest and the cure indicator. In Section \ref{sec:identifiability}, we investigate the identifiability of this model under some regularity conditions. Section \ref{sec:simulation} describes a simulation study, where we investigate the behaviour of this estimator in finite sample. A real data example illustrates the methodology in section \ref{sec:application}. A discussion and some perspectives are given in Section \ref{sec:discussion}.

\section{Model}
\label{sec:model}
\subsection{The model set-up: GEV link function}
Let $(Y_1, S_1, \mathbf X_1, \mathbf Z_1),\ldots, (Y_n, S_n, \mathbf X_n, \mathbf Z_n)$ be independent and identically distributed copies of the random vector $(Y, S, \mathbf X, \mathbf Z)$ defined on the probability space $(\Omega, \mathcal A, \mathbb P)$. For every individual $i=1, \ldots,n$, $Y_i$ is a binary response variable indicating say, the infection status with respect to some disease (that is, $Y_i=1$ if the $i$-th individual is infected, and $Y_i=0$
otherwise), and $S_i$ is a binary variable indicating whether individual $i$ is susceptible to the infection $(S_i=1)$ or immune $(S_i=0)$. If $Y_i=0$, then the value of $S_i$ is unknown. Let $\mathbf X_i=(1, X_{i2}, \ldots, X_{ip})^{\top}$ and $\mathbf Z_i=(1, Z_{i2}, \ldots, Z_{iq})'$ be random vectors of predictors or covariates (both categorical and continuous predictors are allowed). We shall assume in the following that the $\mathbf X_i$'s are related to the infection status, while the $\mathbf Z_i$'s are related to immunity. $\mathbf X_i$ and $\mathbf Z_i$ are allowed to share some components.\\
\cite{wang10} showed that the symmetric link has an inferior performance when the data structure requires a skewed
response probability function. They proposed a link function based on the GEV distribution. The distribution function of $GEV(\mu,\sigma,\tau)$ is given by
\begin{equation}\label{eq0sec2}
G(x|\mu,\sigma,\tau)=
\begin{cases}
\exp\left[-\left\lbrace 1+\tau\frac{x-\mu}{\sigma} \right\rbrace_{+}^{-1/\tau} \right]  & \text{if}~ \tau \neq 0 \\
\exp\left[-\exp\left\lbrace \frac{x-\mu}{\sigma} \right\rbrace  \right]  & \text{if}~\tau =0
\end{cases}
\end{equation}
where $\mu \in \mathbb{R}$ is the location parameter, $\sigma \in \mathbb{R}_{+}$ is the scale parameter, $\tau \in \mathbb{R}$ is the shape parameter and $x_{+}=\max(x,0)$. A more detailed discussion on the extreme value distributions can be found in \cite{c01} and \cite{k00}.\\
Its importance as a link function arises from the fact that the shape parameter $\tau$ purely controls the tail behavior of the distribution.~\\
For a binary response variable $Y_i$ and the vector of explanatory variables $x_i$, let $ \pi(x_i) = \mathbb P(Y_i=1|\mathbf X_i=x_i, S_i)$ the conditional probability of infection. Since we consider the class of Generalized Linear Models, we suggest the GEV cumulative distribution function proposed by \cite{co13} as the response curve given by
%
\begin{equation}\label{eq1sec2}
\begin{split}
\pi(x_i)&= 1-\exp\lbrace\left[(1-\tau(\beta_1+\beta_2 x_{i2}+ \cdots+\beta_p x_{ip}))_{+} \right]^{-1/\tau}  \rbrace \\
&= 1- GEV(-x'_{i}\beta;\tau)
\end{split}
\end{equation}
%
if $\{S_i=1\}$, and by
%
\begin{equation}\label{eq2sec2}
\mathbb P(Y=1|\mathbf X_i=x_i, S_i)=0
\end{equation}
if $\{S_i=0\}$, where $\beta=(\beta_1, \ldots, \beta_p)' \in\mathbb R^p$ is an unknown regression parameter measuring the association between potential predictors and the risk of infection (for a susceptible individual) and $GEV(x;\tau)$ represents the cumulative probability at $x$ for the GEV distribution with a location parameter $\mu=0$, a scale parameter $\sigma=1$, an unknown shape parameter $\tau$. 
~\\~\\
For $\tau \rightarrow 0$, the previous model (\ref{eq1sec2}) becomes the response curve of the log-log model, for $\tau > 0$ and $\tau < 0$ it becomes the Frechet and Weibull response curve respectively, a particular case of the GEV one.~\\
The link function of the GEV model is given by
\begin{equation}\label{eq3sec2}
\frac{1-\left[\log(1-\pi(x_i)) \right]^{-\tau} }{\tau}=x'_{i}\beta=\eta(x_i)
\end{equation}
that represents a noncanonical link function.~\\
Note the parameters $\mu$ and $\sigma$ are set to fixed constants for model identifiability. \cite{wang10} showed that the GEV link model specified in \ref{eq1sec2} is negatively skewed for $\tau < \log(2)-1$ and positively skewed for $\tau > \log(2)-1$. The link function is approximately symmetric at $\tau = \log(2)-1$

\subsection{The proposed estimation procedure} 

A model for the immunity status is defined through the conditional probability $\mathbb P(S=1|\mathbf Z_i)$ of being susceptible to the infection. A common choice for this is the logistic model (see, for example, \cite{fang05} and \cite{lu2008, lu2010} who considered estimation in various survival regression models with a cure fraction):
%
\begin{equation}\label{eq4sec2}
\alpha(z_i)=\log\left(\frac{\mathbb P(S=1|\mathbf Z_i=z_i)}{1-\mathbb P(S=1|\mathbf Z_i=z_i)}\right)=\theta_1+\theta_2 z_{i2}+ \cdots+\theta_q z_{iq}:=z'_{i}\theta
\end{equation}
where $\theta=(\theta_1, \ldots, \theta_q)^{\top}\in\mathbb R^q$ is an unknown regression parameter.

%
\begin{remark}\label{rem1}
\normalfont
We note that the model defined by (\ref{eq1sec2})-(\ref{eq2sec2})-(\ref{eq4sec2}) can be viewed as a zero-inflated Bernoulli regression model, with generalized extreme value link for the binary response of interest and logit link for the zero-inflation component. As far as we know, no theoretical investigation of this model has been undertaken yet. Such a work is carried out in the following.
\end{remark}

\noindent
From (\ref{eq1sec2}), (\ref{eq2sec2}), and (\ref{eq4sec2}), a straightforward calculation yields that
\begin{equation}
\begin{split}
\mathbb{P}(Y=1|\mathbf X_i=x_i, \mathbf Z_i=z_i)&= \pi(x_i)\times \alpha(z_i) \\
&=(1-GEV(-x'_i\beta;\tau))\times\frac{e^{z'_i\theta}}{1+e^{z'_i\theta}}
\end{split}
\end{equation}
Let $\psi:=(\beta',\theta',\tau)'$ denote the unknown $k$-dimensional ($k=p+q+1$) parameter in the conditional distribution of $Y$ given $\mathbf X_i$ and $\mathbf Z_i$. $\psi$ includes both $\beta$ (considered as the parameter of interest) and $\theta$ (considered as a nuisance parameter). Now, the likelihood for $\psi$ from the independent sample $(Y_i, S_i, \mathbf X_i, \mathbf Z_i)$ $(i=1,\ldots,n)$ (where $S_i$ is unknown when $Y_i=0$) is as follows:
\scriptsize{
\begin{equation}\label{eq5sec2}
L_n(\psi)=\prod_{i=1}^n \left\{\left[ \left(1-\exp\lbrace\left[(1+\tau\beta'X_i)_{+} \right]^{-1/\tau}\right) \times
\frac{e^{\theta'Z_i}}{1+e^{\theta'Z_i}}\right] ^{Y_i}\left[ 1- \left( 1-\exp\lbrace\left[(1+\tau\beta'X_i)_{+} \right]^{-1/\tau}\right)\times
\frac{e^{\theta'Z_i}}{1+e^{\theta'Z_i}}\right] ^{1-Y_i}
\right\}.
\end{equation}
}
\normalsize
We define the maximum likelihood estimator $\widehat\psi_n:=(\hat \beta'_{n},\hat\theta'_{n},\hat{\tau}_n)$ of $\psi$ as the solution (if it exists) of the $k$-dimensional score equation
%
\begin{equation}\label{eq6sec2}
\dot l_n(\psi)=\frac{\partial l_n(\psi)}{\partial\psi}=0,
\end{equation}
where $l_n(\psi):=\log L_n(\psi)$ is the log-likelihood function. 

\section{Identifiability}
\label{sec:identifiability}
We first state some regularity conditions that will be needed to ensure identifiability.
\begin{description}
\item[C1] The covariates are bounded that is, there exist compact sets $F\subset\mathbb R^p$ and $G\subset\mathbb R^q$ such that $\mathbf X_i\in F$ and $\mathbf Z_i\in G$ for every $i=1,2,\ldots$ For every $i=1,2,\ldots$, $j=2,\ldots,p$, $k=2,\ldots,q$, $\mbox{var}[X_{ij}]>0$ and $\mbox{var}[Z_{ik}]>0$. For every $i=1,2,\ldots$, the $X_{ij}$
$(j=1,\ldots,p)$ are linearly independent, and the $Z_{ik}$ $(k=1,\ldots,q)$ are linearly independent.
\item[C2] Let $(\beta_{0}^{'}, \theta_{0}^{'})^{'}$ denote the true parameter value. $\beta_0$ and $\theta_0$ lie in the interior of known compact sets $\mathcal B \subset\mathbb R^p$ and $\mathcal G\subset\mathbb R^q$ respectively.
%
%
\item[C3] There exists a continuous covariate $V$ which is in $\mathbf Z$ but not in $\mathbf X$ that is, if $\beta_V$ and $\theta_V$ denote the coefficients of $V$ in the linear predictors (\ref{eq1sec2}) and (\ref{eq4sec2}) respectively, then $\beta_V =0$ and $\theta_V \neq 0$. At a model-building stage, it is known that $V$ is in $\mathbf Z$.
\end{description}
The conditions \textbf{C1} and \textbf{C2} are classical conditions for identifiability and asymptotic results in standard regression models (see, for example, \cite{gm81} and \cite{g01}). The condition \textbf{C3}, which imposes some restrictions on the covariates, is required for identifiability of $\beta$ and $\theta$ in the joint model (\ref{eq1sec2})-(\ref{eq2sec2})-(\ref{eq4sec2}) (we may alternatively assume that the continuous covariate $V$ is in $\mathbf X$ but not in $\mathbf Z$). In the following, we will assume that $V$ is in $\mathbf Z$ but not in $\mathbf X$, with $\theta_V:=\theta_l$ for some $l\in\{2,\ldots,p\}$, and for the $i$-th individual, we will denote $V_i$ by $Z_{il}$. Note \cite{ddd11} use a similary condition to prove the identifiability of the mixture model with logit link function for both models.
We are now in position to prove the following result:

\begin{theorem}[Identifiability]\label{th1}
Under the conditions \textbf{C1}-\textbf{C2}-\textbf{C3}, the model (\ref{eq1sec2})-(\ref{eq2sec2})-(\ref{eq4sec2}) is identifiable; that is, $L_1(\beta,\theta,\tau)=L_1(\tilde{\beta},\tilde{\theta},\tilde{\tau})$ almost surely implies $\beta=\tilde{\beta}$, $\theta=\tilde{\theta}$  and $\tau=\tilde{\tau}$.
\end{theorem}

\noindent
\textbf{Proof of Theorem \ref{th1}} 
~\\
\normalfont
 Suppose that $L_1(\beta,\theta,\tau)=L_1(\tilde{\beta},\tilde{\theta},\tilde{\tau})$ almost surely. Under \textbf{C1} and \textbf{C2}, there exists a positive constant $c_1$ such that for every $x \in F$, $z \in G$, $c_1<\mathbb P(Y=1|x,z)<1-c_1$. Thus we can find a $\omega\in \Omega$, outside the negligible set where $L_1(\beta,\theta,\tau) \neq L_1(\tilde{\beta},\tilde{\theta},\tilde{\tau})$, and such that $Y(\omega)=1$ when $\mathbf X = x$ and $\mathbf Z= z$. For this $\omega$, $L_1(\beta,\theta,\tau)=L_1(\tilde{\beta},\tilde{\theta},\tilde{\tau})$ becomes
\begin{equation*}
\left(1-\exp\lbrace\left[(1-\tau\beta'x)_{+} \right]^{-1/\tau}\right) \times
\frac{e^{\theta'z}}{1+e^{\theta'z}} = \left(1-\exp\lbrace\left[(1-\tilde{\tau}\tilde{\beta}'x)_{+} \right]^{-1/\tilde{\tau}}\right) \times
\frac{e^{\tilde{\theta}'z}}{1+e^{\tilde{\theta}'z}} 
\end{equation*}
~\\
This can be rewritten as
%
\begin{equation}\label{idt1}
\frac{1+e^{-\tilde{\theta}z}}{1+e^{-\theta z}}=\frac{1-\exp\left( \left[(1+\tilde{\tau}\tilde{\beta}'x)_{+} \right]^{-1/\tilde{\tau}}\right) }{1+\exp\left( \left[ \tau\beta'x)_{+} \right]^{-1/\tau}\right) }
\end{equation}
~\\
Now, under the condition \textbf{C3}, taking the partial derivative of both sides of (\ref{idt1}) with respect to the $k$-th component of $z$ ($Z_{ik}$ is a continuous covariate) yields
\begin{equation*}
\frac{-\tilde{\theta}_{k}e^{-\tilde{\theta}'z}\left(1+e^{-\theta' z} \right) + \theta_k e^{-\theta' z}\left(1+e^{-\tilde{\theta}'z} \right)}{\left( 1+e^{-\theta'z}\right)^2}=0
\end{equation*}
It follows that
\begin{equation*}
\frac{\theta_k e^{-\theta' z}}{\tilde{\theta}_k e^{-\tilde{\theta}'z}}=\frac{1+e^{-\theta'z}}{1+e^{-\tilde{\theta}'z}}
\end{equation*}

It follows that
\begin{equation*}
\frac{\theta_k}{\tilde{\theta}_k}=\frac{1+e^{\theta'z}}{1+e^{\tilde{\theta}'z}} ~\Leftrightarrow ~
\theta_k (1+e^{\tilde{\theta}'z})=\tilde{\theta}_k (1+e^{\theta'z})
\end{equation*}
Differentiating both sides of this equality with respect to the $l$-th component of $z$ further yields $(\theta-\tilde{\theta})'z=0$, which implies that $\theta=\tilde{\theta}$ under \textbf{C1}. It remains to show that $\beta = \tilde{\beta}$ and $\tau = \tilde{\tau}$, which reduces to the identifiability problem in the standard GEV regression model. We have that $\beta = \tilde{\beta}$ and $\tau = \tilde{\tau}$ under \textbf{C1} (see \cite{wang10} for example), which concludes the proof.

\section{Simulation study}
\label{sec:simulation}
\subsection{Study design}
In this section, we investigate the numerical properties of the maximum likelihood estimator $\hat{\beta}_n$, under various conditions. The simulation setting is as follows. We consider the following models for the infection status:
\begin{equation}\label{simeq1}
\begin{cases}
\mathbb{P}(Y_i =1|X_i, S_i)=1-GEV(\beta_{1}X_{i1} +\beta_{2}X_{i2}+\beta_{i3}X_{i3};\tau) & \text{if}~S_i =1 \\
\mathbb{P}(Y_i =1|X_i, S_i)=0 & \text{if}~S_i =0
\end{cases}
\end{equation}
and the immunity status:
\begin{equation}\label{simeq2}
\textit{logit}(\alpha(Z_i))=\log\left( \frac{\mathbb{P}(S_i =1| Z_i)}{1-\mathbb{P}(S_i =1| Z_i)}\right)=\theta_{1}Z_{i1} +\theta_{2}Z_{i2}+\theta_{i3}Z_{i3}
\end{equation}
where $X_{i1}=Z_{i1}=1$ for each individual $i$ $(i=1,...,n)$. The covariates $X_{i2}$ and $X_{i3}$ are independently drawn from normal $\mathcal{N}(0,1)$, normal $\mathcal{N}(0,1)$ respectively.\\
The true parameter $\beta$ is set such that the proportion of 1’s in the simulated data sets is around $10\%$ (considered as Model $\mathcal{M}_1$: $\beta = (-2.1,1.2,0)'$) and $25\%$ (considered as Model $\mathcal{M}_2$, $\beta = (-1.3,0,2.5)'$). With the same value of $\beta$ 
we carry out our studies under two scenarios based on the percentages of immunes in the sample as follows:
\begin{description}
\item[$\bullet$]\textbf{Scenario 1}: The regression parameter $\theta$ is chosen as $\theta=(0.85,-1.8,0.5)$. In this setting, the average proportion of immunes data is $30\%$.
\item[$\bullet$]\textbf{Scenario 2}: The regression parameter $\theta$ is chosen as $\theta=(0.2,1.5,-1.71)$. In this setting, the average proportion of immunes data is $70\%$.
\end{description}
An i.i.d. sample of size $n\geq 1$ of the vector $(Y,S,X,Z)$ is generated from the model (\ref{simeq1}-\ref{simeq2}), and for each individual $i$, we get a realization $(y_i,s_i,x_i,z_i)$, where $s_i$ is considered as unknown if $y_i =0$. A maximum likelihood estimator $\hat{\beta}_n$ of $\beta = (\beta_{1},\beta_{2},\beta_{3})'$ is obtained from this incomplete dataset by solving the score equation (\ref{eq6sec2}), using the \textbf{\texttt{optim}} function of the software \textbf{\texttt{R}}. An estimate is also obtained for $\theta = (\theta_{1},\theta_{2},\theta_{3})'$, but $\theta$ is not the primary parameter of interest hence we only focus on the simulation results for $\hat{\beta}_n$. The finite-sample behavior of the maximum likelihood estimator $\hat{\beta}_n$ was assessed for several sample sizes $(n = 100,500,1000,1500)$ based on the two scenarios.
\subsection{Results}
For each configuration (sample size, percentage of immunes) of the design parameters, $N = 2000$ samples were obtained. Based on these $N=2000$ replicates, we obtain averaged values for the estimates of the parameters $\beta_j$, $j=1,\ldots,5$, which are calculated as $N^{-1}\sum_{k=1}^{N}\hat{\beta}_{j,n}^{(k)}$, where $\hat{\beta}_{j,n}^{(k)}$ is the estimate
obtained from the $k$-th simulated sample. The quality of estimates is evaluated by using the Bias and the Root Mean Square Error (RMSE) defined as, for $j=1,2,3$
\begin{equation*}
\begin{split}
\text{Bias}(\hat{\beta}_{n,j}) &= \mathbb{E}(\hat{\beta}_{n,j} -\beta)~ \approx ~\frac{1}{N}\sum_{k=1}^{N}\left( \hat{\beta}_{j,n}^{(k)} -\beta\right) \\
\text{RMSE}(\hat{\beta}_{n,j}) &= \sqrt{\mathbb{E}\left[ (\hat{\beta}_{n,j} -\beta)^2\right]}~ \approx ~\sqrt{\frac{1}{N}\sum_{k=1}^{N}\left( \hat{\beta}_{j,n}^{(k)} -\beta\right) ^2}
\end{split}
\end{equation*}
The results from the model (\ref{simeq1}-\ref{simeq2}) are summarized in Table \ref{tab1} and \ref{tab2}. We compare these results to the ones obtained from a "naive" method where: i) we consider every individual $i$ such that $\{Y_i = 0\}$ as being susceptible but uninfected, that is we ignore the possible immunity of this individual, ii) we apply usuals generalized extreme value regression and logistic regression analysis to the resulting dataset. The results of such "naive" analysis for model $\mathcal{M}_1$ are given in Table \ref{tab3} (the results for models $\mathcal{M}_2$ yield similar observations and thus, they are not given here).
%
%
\begin{table}[h]
\tabcolsep=4pt
\def\arraystretch{1.05}
\caption{Simulation results for Model $\mathbf{\mathcal{M}_1}$: $\mathbf{\beta = (-2.1,1.2,0)'}$)}
\label{tab1}
\begin{scriptsize}
\begin{center}
\begin{tabular}{cccccccccc}
\hline
&&& \multicolumn{3}{c}{30\% of immune} && \multicolumn{3}{c}{70\% of immune}\\
\cline{4-6} \cline{8-10}
n& && $\widehat\beta_{1,n}$ & $\widehat\beta_{2,n}$ & $\widehat\beta_{3,n}$ && $\widehat\beta_{1,n}$ & $\widehat\beta_{2,n}$ & $\widehat\beta_{3,n}$ 
\\\hline
100&MLE && -2.368& 1.509 & 0.013&& -2.551 & 1.659 &0.211 \\
&BIAS && -0.268 & 0.309 & 0.013&& -0.451 & 0.459 & 0.211  \\
&RMSE &&  1.652& 0.941 & 0.303&&  1.806& 1.627 &1.995  \\
\hline
500&MLE && -2.033& 1.270 &0.001 && -2.289& 1.494& -0.001  \\
&BIAS && 0.067 & 0.071 &0.001 && 0.189 & 0.294 & 0.001  \\
&RMSE && 0.642 & 0.593 & 0.060&& 1.340 & 1.085 & 0.142 \\
\hline
1000&MLE &&-2.081 & 1.232 & 0.000&& -2.075 & 1.299 & 0.001\\
&BIAS && 0.019 & 0.032 & 0.000&& 0.025 & 0.099 & 0.001  \\
&RMSE && 0.463 & 0.393 &0.036 && 0.768 & 0.696 & 0.071 \\
\hline
1500& MLE&& -2.085& 1.223 & 0.000&& -2.080 & 1.256 & 0.000\\
&BIAS && 0.015 & 0.023 & 0.000&&  0.019& 0.056 & 0.000  \\
&RMSE && 0.390 & 0.323 & 0.029&& 0.656& 0.588 & 0.055 \\
\hline
\end{tabular}
\end{center}
\end{scriptsize}
%
\end{table}
%
%
\begin{table}[h]
\tabcolsep=4pt
\def\arraystretch{1.05}
\caption{Simulation results for Model $\mathbf{\mathcal{M}_2}$: $\mathbf{\beta=(-1.3,0,2.5)'}$}
\label{tab2}
\begin{scriptsize}
\begin{center}
\begin{tabular}{cccccccccc}
\hline
&&& \multicolumn{3}{c}{30\% of immune} && \multicolumn{3}{c}{70\% of immune}\\
\cline{4-6} \cline{8-10}
n &&& $\widehat\beta_{1,n}$ & $\widehat\beta_{2,n}$ & $\widehat\beta_{3,n}$ && $\widehat\beta_{1,n}$ & $\widehat\beta_{2,n}$ & $\widehat\beta_{3,n}$ 
\\\hline
100& MLE&&-1.778 &-0.003  & 3.101&& -1.997 & 0.021 & 3.209 \\
&BIAS && -0.478 & -0.003 & 0.601&& -0.697 & 0.021 & 0.709   \\
&\text{RMSE} && 1.680 & 0.859 & 1.117 && 1.684 & 1.410 & 1.260  \\
\hline
500 &MLE&&-1.370 & 0.002 & 2.743&& -1.374&-0.019 &2.822 \\
&BIAS && -0.070 & 0.002 &0.243 && -0.075 & -0.019 &0.322   \\
&RMSE && 0.617 & 0.217 &0.974 && 0.816 & 0.424 & 1.190 \\
\hline
1000 &MLE&& -1.310& 0.001 &2.608 && -1.372& 0.004 &2.639 \\
&BIAS && -0.010 &0.001  & 0.108&&  -0.072&0.004  & 0.139  \\
&RMSE && 0.362 & 0.128 & 0.778&& 0.732 &0.293  & 1.055 \\
\hline
1500&MLE &&-1.300 & 0.000 &2.552 && -1.312&  0.002 & 2.569\\
&BIAS && 0.000 & 0.000 & 0.052&& -0.012&  0.002 & 0.069  \\
&RMSE &&0.270& 0.104   & 0.570&& 0.557& 0.209  & 0.983 \\
\hline
\end{tabular}
\end{center}
\end{scriptsize}
%
\end{table}
%
\newpage
\begin{table}[h!]
\tabcolsep=4pt
\def\arraystretch{1.05}
\caption{Simulation results for Naive analysis of Model $\mathbf{\mathcal{M}_1}$: $\mathbf{\beta = (-2.1,1.2,0)'}$)}
\label{tab3}
\begin{scriptsize}
\begin{center}
\begin{tabular}{cccccccccc}
\hline
&&& \multicolumn{3}{c}{30\% of immune} && \multicolumn{3}{c}{70\% of immune}\\
\cline{4-6} \cline{8-10}
n& && $\widehat\beta_{1,n}$ & $\widehat\beta_{2,n}$ & $\widehat\beta_{3,n}$ && $\widehat\beta_{1,n}$ & $\widehat\beta_{2,n}$ & $\widehat\beta_{3,n}$ 
\\\hline
100&MLE && [-3.526] & [2.487]  & [0.243] && [-6.754] & [5.931] & [0.485] \\
 & && (-5.503) & (3.317) & (0.174) && (-8.818) & (5.914) & (0.502) \\
&BIAS &&[-1.427] & [1.287]& [0.243]&& [-4.654]& [4.730] &[0.485]  \\
 & && (-3.403) & (2.117) & (0.174) && (-6.718) & (4.714) & (0.502) \\
&RMSE && [24.059] & [20.583] & [9.524]&& [27.031] & [36.579] & [20.474]\\
 & && (19.081) & (16.018) & (6.537) && (22.771) & ( 28.923) & (0.502) \\
\hline
500&MLE && [-2.116]& [1.328] & [0.001]&& [-2.391] & [1.005] & [0.001]\\
 & && (-4.096) & (2.249) & (0.001) && (-4.743) & (1.791) & (0.001) \\
&BIAS && [-0.016]& [0.128] & [0.001]&& [-0.292]& [-0.194] & [0.001] \\
 & && (-1.996) & (1.049) & (0.001) && (-2.643) & (0.591) & (0.001) \\
&RMSE && [0.262] & [0.251] & [0.064]&& [0.401] & [0.269] &[0.081] \\
 & && (2.033) & (1.092) & (0.112) && (2.690) & (0.671) & (0.153) \\
\hline
1000&MLE &&[-2.068] & [1.299]  & [0.001]&& [-2.326] & [0.976] & [0.001]\\
 & && (-4.056) & (2.223) & (0.001) && (-4.637) & (1.749) & (0.001) \\
&BIAS &&[0.031] & [0.099] &[0.001] && [-0.226]& [-0.224] & [0.001] \\
 & && (-1.957) & (1.024) & (0.001) && (-2.538) & (0.549) & (0.001) \\
&RMSE && [0.188] & [0.188] &[0.043] &&[0.289]  & [0.256]& [0.053]\\
 & && (1.973) & (1.044) & (0.074) && (2.559) & (0.588) & (0.099) \\
\hline
1500& MLE&&[-2.065] & [1.302] & [0.001]&& [-2.317] & [0.968] & [0.001]\\
 & && (-4.042) & (2.215) & (0.001) && (-4.617) & (1.733) & (0.001) \\
&BIAS && [0.035]& [0.102] & [0.001]&& [-0.217] & [-0.232] & [0.001] \\
 & && (-1.943) & (1.015) & (0.001) && (-2.517) & (0.533) & (0.001) \\
&RMSE && [0.158] &[0.171]  &[0.037] && [0.259] & [0.254] & [0.043]\\
 & && (1.953) & (1.028) & (0.061) && (2.529) & (0.559) & (0.080) \\
\hline
\end{tabular}
\parbox{12.1cm}{
\underline{Note}: \noindent$n$: sample size. $[\cdot]$: Naive analysis with GEV link function. $(\cdot)$: Naive analysis with logit link function.}
\end{center}
\end{scriptsize}
%
\end{table}
From the Table \ref{tab3}, it appears that ignoring the immunity present in the sample results in strongly biased estimates of $\beta$. The bias and the RMSE of the parameters estimate increase with the immune proportion. This results a wrong interpretation of the relationship between the covariate $\mathbf X_2$ and $\mathbf X_3$ and the binary response $Y$.\\ 
From the Tables \ref{tab1} and \ref{tab2}, it appears that the proposed maximum likelihood estimator $\widehat \beta_n$ provides a reasonable approximation of the true parameter value, even when the percentage of immunes is high. While the bias of $\widehat \beta_n$ stays limited, its variability increases with the immune fraction, sometimes drastically when the sample size is small. Consequently, when the sample size is small ($n = 100$) and/or the immune proportion is very high (70\%), the power of the Wald test for nullity of the regression coefficients can be low, compared to the case where there are no immunes. But we note that for moderately large to large sample sizes ($n \geq 500$), the dispersion indicators indicate good performance of the maximum likelihood estimate, even when the immune proportion is up to 50\%. ~\\
Finally, these results indicate that a reliable statistical inference on the regression effects and probabilities of event in the regression model for binary data with a GEV link function and a cure fraction should be based on a sample having, at least, a moderately large size ($n\geq 500$, say) when the immune fraction is low ($n\leq 30\%$), or a large size ($n\geq 1000$) when the immunity attains an average level (about 50\%) of the sample. When the immune proportion is very large (about 70\%), the results should be considered carefully, considering the increase in the variability of the estimates and the skewness of their distributions.

\section{Real data application: study of dengue cases}
\label{sec:application}
In this setion, we consider a study of dengue fever, which is a mosquitoborne viral human disease. A dengue infection confers a partial and transient immunity against a subsequent infection (see \cite{dussart11}). We consider
here a database of size $n = 515$ (with 15.5\% of 1's), which was constituted with individuals recruited in Cambodia, Vietnam, French Guiana, and Brazil (\cite{dussart11}). Each individual $i$ was diagnosed for dengue infection and coded as $Y_i = 1$ if
infection was present and $0$ otherwise. Note that if $Y_i = 0$, then the $i$-th individual may either be immune at the time of analysis (due to a temporary immunity acquired following a previous infection) or susceptible to dengue infection,
albeit not infected. We aim at estimating the risk of infection for those individuals, based on this data set which also includes the following covariates: \textbf{\textit{Age}} and \textit{Weight} (continuous bounded covariates). We first ran a standard logistic regression model of the model defined as follows $\mathcal{M}_0$:
\begin{equation*}
\mathbb{P}(Y=1 | \text{\textbf{\textit{Age}}},\text{\textbf{\textit{Weight}}})= \frac{e^{\beta_1 +\beta_2\times \text{\textbf{\textit{Age}}} + \beta_3\times \text{\textbf{\textit{Weight}}}}}{1+e^{\beta_1 +\beta_2\times \text{\textbf{\textit{Age}}} + \beta_3\times \text{\textbf{\textit{Weight}}}}}. 
\end{equation*}
Then we run a generalized extreme value regression analysis of the model defined as follows $\mathcal{M}$odel 1:
\begin{equation*}
\mathbb{P}(Y_i =1 | \text{\textbf{\textit{Age}}},\text{\textbf{\textit{Weight}}}) = 1- GEV(-(\beta_1 +\beta_2\times \text{\textbf{\textit{Age}}} + \beta_3\times \text{\textbf{\textit{Weight}}};\tau).
\end{equation*}
Then, we estimated the parameters using the model (\ref{eq1sec2})-(\ref{eq2sec2})-(\ref{eq4sec2}) ($\mathcal{M}$odel 2) with:
\begin{equation*}
\mathbb{P}(Y_i =1 | \text{\textbf{\textit{Age}}},\text{\textbf{\textit{Weight}}}) = 1- GEV(-(\beta_1 +\beta_2\times \text{\textbf{\textit{Age}}} + \beta_3\times \text{\textbf{\textit{Weight}}};\tau).
\end{equation*}
and 
\begin{equation*}
\mathbb{P}(S=1 | \text{\textbf{\textit{Age}}})= \frac{e^{\theta_1 +\theta_2\times \text{\textbf{\textit{Age}}}}}{1+e^{\theta_1 +\theta_2\times \text{\textbf{\textit{Age}}}}}. 
\end{equation*}
Note first that the eventual immunity imparted by a past infection is only transient, thus there is no reason why an older individual (who has therefore been exposed longer to the risk of dengue fever) would have a greater probability
of being immune than a younger one. In fact, individual susceptibility to the dengue infection may rather depend on whether the individual benefits or not from some preventive and control measures (such as the application of insecticides to larval habitats in his area, or appropriate water storage and waste disposal practices). Such informations are not available in our dataset. 
\begin{remark}
Selection of regressors for inclusion in both models requires some care. Indeed, it was previously observed in various other zero-inflated regression models that including all available regressors in both count and zero-inflation probabilities can yield lack of identification of model parameters. See for example \cite{ddd11} and \cite{staub13}, who suggest to solve this issue by letting at least one of the covariates included in the bernouilli model to be excluded from
the zero-inflation model (or the converse). For this purpose, the covariate \textit{Weight} was therefore taken as the variable $V$ in condition \textbf{C3} for ZI-GEV model identifiability.
\end{remark}
Since the Wald-type tests of "$\beta_2 = 0$" and "$\theta_2 = 0$" were not significant, we removed the covariate \textit{Age} from models $\mathcal{M}_0$, $\mathcal{M}_2$ and the model for susceptibility ($\mathcal{M}_1$), resulting in a constant proportion of immunes.\\
The fitted model for susceptibility produced the following estimate for the probability of being immune: 
\begin{equation*}
\mathbb{P}(S=0) = 1 - \frac{\exp (-0.3667)}{1+\exp (-0.3667)} \approx 0.591.
\end{equation*}
The final results (only the significant covariates for both models using the Wald testing) of these fitting procedures are given in Table \ref{tab4}.
\begin{table}[h!]
\tabcolsep=4pt
\def\arraystretch{1.05}
\caption{Dengue fever data analysis}
\label{tab4}
%
\begin{center}
\begin{tabular}{ccccccccc}
\hline
& \multicolumn{2}{c}{$\mathcal{M}_0$}&&\multicolumn{2}{c}{$\mathcal{M}_1$}&&\multicolumn{2}{c}{$\mathcal{M}_2$} \\
\cline{2-3}  \cline{5-6} \cline{8-9}
Parameter & Estimate & SE && Estimate & SE && Estimate & SE 
\\\hline
\textit{Intercept} ($\beta_1$) & -0.8201& 0.3637&& -0.2693& 0.1878 &&1.5379 &0.0046  \\
\textit{Weight} ($\beta_2$) & -0.0183&0.0074 && -0.0084 & 0.0037 && -0.1003 &0.0006  \\
\textit{Intercept} ($\theta_1$) & & &&  & && -0.3667&0.1659  \\\\
$\tau$ &  & && &  &&4.278 &0.0427  \\\hline
\\
\text{AIC} & 442.730 & && 443.587 & &&\textbf{442.214} & \\
\hline
\end{tabular}
\parbox{12.1cm}{ ~\\
\underline{Note}: \noindent SE: standard error.}
\end{center}
\end{table}
In the "naive analysis" every uninfected individual is considered as susceptible. We obtain for the model (\ref{eq1sec2})-(\ref{eq2sec2})-(\ref{eq4sec2}) the equations given by:
\begin{equation*}
\mathbb{P}(Y =1 | \text{\textbf{\textit{Weight}}}, S=1) = 1- GEV(-(1.5379 - 0.1003 \times \text{\textbf{\textit{Weight}}});4.278)
\end{equation*}
and 
\begin{equation*}
\mathbb{P}(S =1) = \frac{\exp(-0.3667)}{1+\exp(-0.3667)}.
\end{equation*} 
A closer look at the results from the both standard GEV and logistic regression models and our ZI-GEV regression model reveals some difference in the estimation of the covariates effects on the risk infection of Dengue. Note that, as expected, the risk of infection is overestimated by the naive analysis (standard GEV and logistic regression models) that does not take account of the possible immunity compared to our approch. For example, the probabilities of infection for individuals with weights of 15 kg and 50 kg respectively, are estimated by 0.645 and 0.409 (standard generalized extreme value regression) and 0.264 and 0.167 (from our approach). It is expected that underweighted subjects (those considered to be under a healthy weight) will have higher risks of infection. While both approaches provide the same qualitative conclusions: the probability of dengue infection is higher for individuals in case of underweight (caused by malnutrition for example), they differ on their estimations of the risk of infection. Our approach takes account of the possible immunity imparted by a past infection and therefore, it is reasonable to think that the resulting estimations of the infection probabilities provide a more realistic picture of the infection risk for this data set. In particular, the estimates provided by our approach suggest that underweight constitutes a major risk factor for dengue infection, irrespectively of age.

\section{Discussion and perspectives}
\label{sec:discussion}
In this paper, we have considered the problem of estimating the regression model with a generalized extreme value link function from a sample of binary response data with a cure fraction. The estimator we propose is obtained by maximizing a likelihood function, which is derived from a joint regression model for the binary response of interest and
the cure indicator, considered as a random variable whose distribution is modeled by a logistic regression (the proposed joint model can thus be viewed as a zero-inflated Generalized extreme value regression model, with GEV link for the binary response of interest and logit link for the zero-inflation component). we have estabilished the identifiability of the proposed model and investigated its finite-sample properties via simulations.~\\
Several open questions now deserve attention. Study of existence, consistency, and asymptotic normality of this proposed estimator. We can compare these results to the ones obtained from a "naive" method where: i) we consider every individual i such that ${Y_i = 0}$ as being susceptible but uninfected, that is we ignore the eventual immunity of this individual, ii) We apply a usual GEV regression analysis to the resulting dataset.~\\
In regression analysis for binary data, it is usually of interest to estimate the probability of infection $\pi(\mathbf x)=\mathbb P(Y=1|\mathbf X=\mathbf x)$, for some given value $\mathbf x$ of the covariates and to investigate its properties. Another issue of interest deals with the inference in the generalized extreme value regression model with a cure fraction, in a high-dimensional setting, when the predictor dimension is much larger than the sample size (this problem arises, for example, in genetic studies where high-dimensional data are generated using microarray technologies).

\bibliographystyle{Chicago}

\end{document}